\documentclass[useAMS,usenatbib]{mn2e} 
\usepackage{lscape}
\usepackage{graphicx} 
\usepackage{amssymb} 
\usepackage[usenames]{color}

\title[Am stars spectroscopy with CAOS]{CAOS spectroscopy of Am stars {\it Kepler} targets\thanks{Based 
          on observations made with the Catania Astrophysical Observatory Spectropolarimeter (CAOS) operated by the Catania Astrophysical
          Observatory}} 
\author[G. Catanzaro et al.]{G. Catanzaro$^{1}$\thanks{E-mail: gca@oact.inaf.it}, V. Ripepi$^{2}$, K. Biazzo$^{1}$, I. Bus\'a$^{1}$,
                             A. Frasca$^{1}$, F. Leone$^{3,1}$, \and M. Giarrusso$^{3,1}$, M. Munari$^{1}$, S. Scuderi$^{1}$ 
\\  
$^{1}$INAF-Osservatorio Astrofisico di Catania, Via S.Sofia 78, I-95123, Catania, Italy\\ 
$^{2}$INAF-Osservatorio Astronomico di Capodimonte, Via Moiariello 16, I-80131, Napoli, Italy\\
$^{3}$Universit\'a degli studi di Catania, Via S.Sofia 78, I-95123, Catania, Italy\\
} 
 
\date{Accepted   Received ; in original form } 
 
\pagerange{\pageref{firstpage}--\pageref{lastpage}} \pubyear{2002} 
 
\def\LaTeX{L\kern-.36em\raise.3ex\hbox{a}\kern-.15em 
 T\kern-.1667em\lower.7ex\hbox{E}\kern-.125emX} 
 
\begin{document} 
 
\label{firstpage} 
 
\maketitle 
 
\begin{abstract} 
The {\it Kepler} space mission and its {\it K2} extension provide photometric time
series data with unprecedented accuracy.  These data challenge our current
understanding of the metallic-lined A stars (Am stars) for what concerns the
onset of pulsations in their atmospheres. It turns out that
the predictions of current diffusion models do not agree with observations.
To understand this discrepancy, it is of crucial importance to obtain 
ground-based spectroscopic observations of Am stars in the {\it Kepler} and {\it K2} 
fields in order to determine the best estimates of the stellar parameters.

In this paper, we present a detailed analysis of high-resolution
spectroscopic data for seven stars previously classified as Am stars.
We determine the effective temperatures, surface gravities, projected 
rotational velocities, microturbulent velocities and chemical abundances
of these stars using spectral synthesis.  These spectra were obtained with
{\it CAOS}, a new instrument recently installed at the observing station of the 
Catania Astrophysical Observatory on Mt. Etna. Three stars have already
been observed during quarters Q0-Q17, namely: HD\,180347, HD\,181206, and HD\,185658,
while HD\,43509 was already observed during {\it K2} C0 campaign.

We confirm that HD\,43509 and HD\,180347 are Am stars, while HD 52403, HD\,50766, 
HD\,58246, HD\,181206 and HD\,185658 are marginal Am stars. By means of non-LTE analysis, 
we derived oxygen abundances from O{\sc I}$\lambda$7771--5{\AA} triplet and we also 
discussed the results obtained with both non-LTE and LTE approaches.

\end{abstract} 
 

\setcounter{equation}{0} 

\begin{table*} 
\scriptsize
\begin{center}
\caption{Main photometric data and physical parameters estimated from SED fitting for the target stars. The different columns show:
 (1) the HD number; (2) the EPIC or KIC identifier; (3) and (4) the adpted $B$ and $V$ magnitudes ($\sigma_B$,$\sigma_V$$\sim$0.020,0.015 mag, 
 respectively); (5) the parallax \citep[][]{leeuwen}; (6) the $E(B-V)$ values (the uncertainty is $\sim$0.025 mag for all
 the stars); (7) the bolometric correction in the $V$ band \citep[after][]{bcp98}; (8-10) the estimated $T_{\rm eff}$, $\log g$ and $[Fe/H]$ from 
 the SED fitting (see the text for a discussion about errors); (11) and (12) the luminosity for two choices of the metal abundance
 Z=0.014 and Z=0.03, respectively (see Sect.~\ref{HR}).}  
\label{tabPhotometry} 
\begin{tabular}{cccccccccccc} 
\hline 
\hline 
\noalign{\medskip} 
HD  & KIC/EPIC &$B$& $V$ &  $\pi$&$E$($B$$-$$V$) & $BC_V$ &$T^{SED}_{\rm eff}$ & $\log g^{SED}$ & $[M/H]^{SED}$  &  $\log L/{\rm  L}_\odot$    &   $\log L/{\rm  L}_\odot$ \\ 
\noalign{\medskip} 
     &   &mag      & mag & mas& mag                         &     mag
     &        K                & cm/s$^{2}$    &   dex     &
     Z=0.014$^{a}$   & Z=0.03$^{b}$ \\ 
(1)   &  (2)    & (3)         & (4)         &    (5)&(6) &(7)&(8) &(9)& (10) & (11) & (12) \\ 
\noalign{\medskip} 
\hline 
\noalign{\smallskip}                                                 
 43509   &   202059336  &  9.185  &  8.907   &                & 0.09~~ &  0.094     &  7500$\pm$250   & 3.50$\pm$0.50 &$-$0.50$\pm$0.50~~ & 1.28$\pm$0.16 & 1.33$\pm$0.16 \\
 50766   &   209907943  &  7.924  &  7.783   &                & 0.075  & $-$0.059~~ &  9000$\pm$250   & 4.50$\pm$0.50 &   0.50$\pm$0.30   &  1.68$\pm$0.14 &  1.72$\pm$0.14\\
 52403   &   209536243  &  7.165  &  7.038   & 6.52$\pm$0.58  & 0.055  &  0.011     &  8750$\pm$250   & 4.50$\pm$0.50 &   0.50$\pm$0.30   &  1.52$\pm$0.08 &  1.52$\pm$0.08\\
 58246$^{c}$  &         &  7.590  &  7.291   & 7.55$\pm$1.48  & 0.050  &  0.121     &  7750$\pm$250   & 4.50$\pm$0.50 &   0.50$\pm$0.30   &  1.24$\pm$0.17 &  1.24$\pm$0.17\\
 180347  &    12253106  &  8.647  &  8.388   &                & 0.06~~ &  0.067     &  7750$\pm$250   & 3.50$\pm$0.50 &   0.20$\pm$0.25   &  1.43$\pm$0.10  &  1.47$\pm$0.10\\
 181206  &     9764965  &  9.074  &  8.831   &                & 0.07~~ &  0.090     &  7750$\pm$250   & 3.50$\pm$0.50 &   0.50$\pm$0.30   &  1.46$\pm$0.11 &  1.50$\pm$0.11\\
 185658  &     9349245  &  8.333  &  8.119   &                & 0.055  &  0.038     &  8250$\pm$250   & 4.50$\pm$0.50 &$-$0.50$\pm$0.50~~ &  1.36$\pm$0.38 &  1.41$\pm$0.37\\
\noalign{\smallskip} 
\hline 
\end{tabular}
\end{center}
\begin{flushleft}
     $^{a}$ Value calculated with Eq.~\ref{eq} except for HD\,52403 and HD\, 58246 (see Sect.~\ref{HR}).\\
     $^{b}$ Value calculated with Eq.~\ref{eq1} except for HD\,52403 and HD\, 58246 (see Sect.~\ref{HR}).\\
     $^{c}$ The EPIC ID for HD\,58246 is missing likely because its position was $\sim$4 degree out of the final K2 C0 field of view.
\end{flushleft}
\end{table*} 

\section{Introduction}

Among the Chemically Peculiar stars of the main sequence, the Am
sub-group shows Ca{\sc ii} K-line too early for their hydrogen line
types, while metallic-lines appear too late, such that the spectral
types inferred from the Ca{\sc ii} K- and metal-lines differ by five
or more spectral subclasses. The marginal Am stars (denoted with Am:)
are those whose difference between Ca{\sc ii} K- and metal-lines is
less than five subclasses.  The commonly used classification for this
class of objects includes three spectral types prefixed with {\it k},
{\it h}, and {\it m}, corresponding to the K-line, hydrogen-lines and
metallic lines, respectively. The typical chemical pattern show
underabundances of Ca and/or Sc and overabundances of the
Fe-peak elements, Y, Ba and of rare earths elements \citep{adelman97,fossati07}.

It is commonly believed that the strength of the metal lines is due to
the interplay between gravitational settling and radiative
acceleration in an A-type star where magnetic field should be weak or
absent. In this scenario, the Am stars should rotate slower than about
120 km s$^{-1}$ to allow radiative diffusion to compete with 
meridional circulation \citep[see, e.g.][and references therein]{Charbonneau1993}.

Most Am stars appear to be members of binary systems with periods
between 2 and 10 days \citep[e.g.][]{Smalley2014}.

One important open question in the framework of Am stars concerns the
presence of pulsations. For many years it was thought that Am stars
did not pulsate, since their He{\sc ii} 
ionization zone should be fully depleted of helium  due to atomic
diffusion, i.e. radiative levitation and, in particular, gravitational
settling, thus preventing the $\kappa$-mechanism to excite a
$\delta$ Scuti type pulsation. This scenario applies to Am stars as they rotate
slowly. On the contrary, normal A stars are usually rapid rotators and remain mixed 
because of turbulence induced by meridional circulation; thus in their interiors 
the pulsations can be excited by the $\kappa$-mechanism \citep{cox79,Turcotte2000}.

This theoretical scenario has been recently questioned by
intensive ground-based \citep[SuperWASP survey,][]{smalley11}
and space-based \citep[{\it Kepler} mission,][]{balona11} observations
which have clearly demonstrated that many Am/Fm stars do pulsate. Before these 
recent surveys, pulsations were observed only in two Am stars, namely HD\,1097 \citep{kurtz89} 
and HD\,13079 \citep{martinez99}. \citet{smalley11}, for
example, found that about 169, 30 and 28 Am stars out of a total of
1600 show $\delta$~Sct, $\gamma$~Dor or Hybrid pulsations, respectively
\citep[see, e.g.,][for a definition of these classes]{griga2010}. These
authors found also that the positions in the Hertzsprung-Russel (HR)
diagram of Am stars pulsating as $\delta$~Sct are confined between the
red and blue radial fundamental edges, and this result has been also
confirmed by \citet{balona11} and \citet{catanzaro13}. 
Pulsating Am stars show oscillations with lower amplitude than
normal abundance $\delta$ Scuti stars. 
There is still not a satisfactory explanation for this amplitude
difference. 

Recently, \citet{Balona2013} suggested that starspots could be the origin of
rotational-type variability found in a significant fraction of A stars, 
whereas \citet{Balona2012,Balona2014} proposed that about 2 percent of the 
A stars exhibit flares. As to Am stars, based on 50-days {\it Kepler } time 
series, \citet{balona11} found suspected rotational modulation in a sample 
of 10 stars. This result has been very recently confirmed by
\citet{Balona15}, who analyzed a sample of 15 Am stars with 4-years of
{\it Kepler} data plus 14 Am stars from the {\it K2} Campaign 0 (C0) field\footnote{http://keplerscience.arc.nasa.gov/K2/Fields.shtml}
finding that most of these objects show a rotational
modulation due to starspots whereas two stars exhibit flares.

In this context, the present paper is a further step in our programme devoted to determine
photospheric abundances in Am stars that lie in the Kepler field of
view, by means of high resolution spectra. Three papers have already
been published on this topics: \citet{balona11}, \citet{catanzaro13},
and \citet{catanzaro14}, in which, for a total of 11 stars,
fundamental astrophysical quantities, such as effective temperatures,
gravities and metallicities have been derived.  Such kind of studies
are crucial in order {\it i)} to put constraints on the processes
occurring at the base of the convection zone in non-magnetic stars and
{\it ii)} to try to define the locus on the HR diagram occupied by
pulsating Am stars.

With these goals in mind, we present a complete analysis of other
seven stars previously classified as Am stars.  Three targets have
already been observed by Kepler during quarters Q0-Q17 (namely, HD\,180347, HD\,181206, and
HD\,185658), while the other five stars had been proposed for
observations in the context of the {\it K2} phase mission. However, only
HD\,43509 was actually targeted during {\it K2} C0 campaign, because this
star was the only one that falled on the silicon when the final C0
field center was decided \citep[see][for details on the {\it Kepler} and
{\it K2} observations for the target stars]{Balona15}. For our purposes 
high-resolution spectroscopy is the best tool principally for two reasons, 
{\it i)} the blanketing due to the chemical peculiarities in the atmospheres 
of Am stars alters photometric colors and then fundamental stellar parameters
based on them may not be accurate \citep[see][]{catanzaro12} and {\it ii)} 
the abnormal abundances coupled with rotational velocity result
in a severe spectral lines blending which makes difficult the
separation of the individual lines. Both problems could be overcome
only by matching synthetic and observed spectra.

\begin{figure}                                       
\includegraphics[width=6.8cm,angle=-90]{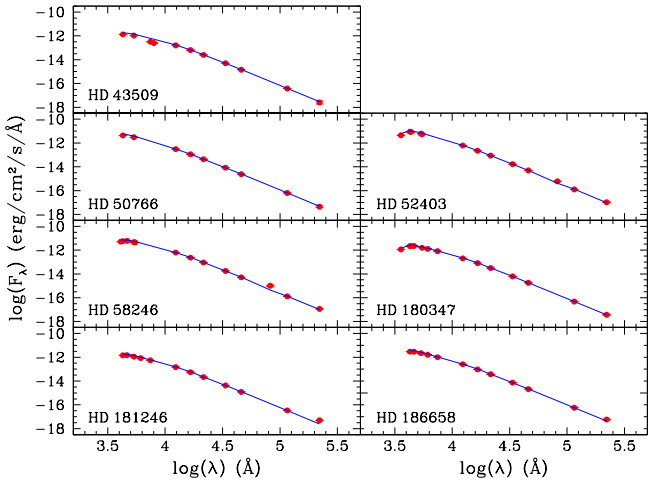}     
\caption{SED for the target stars (filled circles). The solid line
  represents the fit to the data obtained by means of the $VOSA$ tool
  (see text).} 
\label{sed} 
\end{figure} 

\section{Observation and data reduction}
\label{obs}

Spectroscopic observations of our sample of seven stars (see
Table~\ref{tabPhotometry} for the list of targets) were carried out
with the new {\it Catania Astrophysical Observatory Spectropolarimeter} (CAOS) 
which is a fiber fed, high-resolution, cross-dispersed echelle spectrograph 
\citep{leo15,spano06,spano04} installed recently at the cassegrain focus of the
91\,cm telescope of the {\it ``M. G. Fracastoro''} observing station of the 
Catania Astrophysical Observatory (Mt. Etna, Italy).

Our spectra were obtained in 2014, between March and
November. Exposure times have been tuned in order to obtain for the
stars a signal-to-noise ratio of at least 100 in the continuum in the 
390--800~nm, with a resolution of R = 45\,000, as measured from ThAr 
and telluric lines.

The reduction of all spectra, which included the subtraction of the
bias frame, trimming, correcting for the flat-field and the scattered
light, extraction for the orders, and wavelength calibration,
was done using the NOAO/IRAF packages\footnote{IRAF is distributed by
  the National Optical Astronomy Observatory, which is operated by the
  Association of Universities for Research in Astronomy, Inc.}. 
Given the importance of Balmer lines in our analysis, we paid much more attention 
in the normalization of the corresponding spectral orders.
In particular, we divided the spectral order containing H$\beta$ by a pseudo-continuum 
obtained combining the continua of the previous and subsequent echelle orders, as
already outlined by \citet{leo93}. This procedure was not necessary for the H$\alpha$, 
since spectral coverage of the order is wider than the line
itself and it was possible to properly fix the continuum offset.

The IRAF package {\it rvcorrect} was used to determine the barycentric velocity 
correcting the spectra for the Earth's motion.

\begin{table*}
\centering
\caption{Results obtained from the spectroscopic analysis of the
  sample of Am stars presented in this work. The different columns
  show: identification; effective temperatures; gravity ($\log g$); ODF value;
  microturbolent velocity ($\xi$); rotational velocity (v\,$\sin i$).
}
\begin{tabular}{rrcccc}
\hline
\hline
\noalign{\medskip} 
 HD~~  & T$_{\rm eff}$~~~~~~&  $\log g$ &  [ODF] &     $\xi$      &  v $\sin i$ ~~  \\
       &           (K)~~~~~~&           &        & (km s$^{-1}$)~ & (km s$^{-1}$)~~~ \\
\noalign{\medskip} 
\hline    
\noalign{\smallskip} 
43509  &   7900\,$\pm$\,150 & 3.97\,$\pm$\,0.12 & 0.2 & 4.1\,$\pm$\,0.4 & 28\,$\pm$\,1 \\   
50766  &   9000\,$\pm$\,200 & 3.90\,$\pm$\,0.10 & 1.0 & 4.6\,$\pm$\,0.9 & 28\,$\pm$\,1 \\      
52403  &   8500\,$\pm$\,200 & 3.50\,$\pm$\,0.10 & 0.2 & 4.6\,$\pm$\,0.7 & 17\,$\pm$\,1 \\      
58246  &   8000\,$\pm$\,150 & 3.70\,$\pm$\,0.10 & 0.5 & 4.0\,$\pm$\,0.5 & 55\,$\pm$\,2 \\      
180347 &   7900\,$\pm$\,140 & 3.85\,$\pm$\,0.07 & 0.2 & 4.7\,$\pm$\,0.4 & 12\,$\pm$\,1 \\    
181206 &   7800\,$\pm$\,140 & 3.80\,$\pm$\,0.10 & 0.5 & 2.3\,$\pm$\,0.5 & 87\,$\pm$\,3 \\    
185658 &   8300\,$\pm$\,200 & 4.00\,$\pm$\,0.30 & 0.5 & 3.1\,$\pm$\,0.5 & 80\,$\pm$\,3 \\    
\noalign{\smallskip} 
\hline
\end{tabular}
\label{param}
\end{table*}    
     
\begin{figure} 
\begin{center}
\includegraphics[width=13cm]{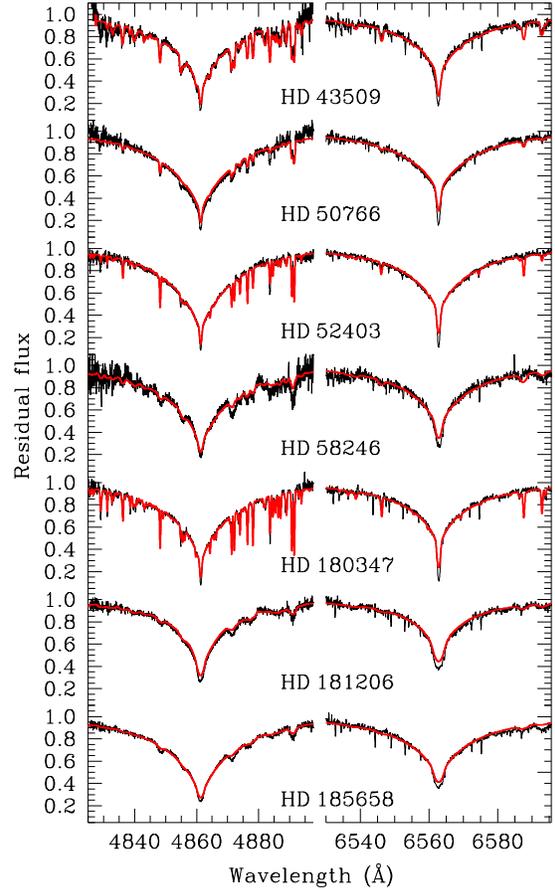}    
\caption{Portions of the spectral echelle orders centered around H$\beta$ and H$\alpha$ for the program stars. 
          Overimposed (red lines) the synthetic spectra computed with the parameters reported in Tab.~\ref{param}.} 
\label{balmer} 
\end{center}   
\end{figure}

\section{Spectral Energy Distributions}
\label{parameters_from_sed} 
With the unique goal of speeding up the spectroscopic investigation of the
target stars, we estimated first guess stellar parameters by
comparing the observed Spectral Energy Distributions (SEDs) with selected model atmosphere. 
To this aim we  adopted the $VOSA$ tool \citep{Bayo2008}. The first step was to
collect the photometric data by means of the $VOSA$ package
itself. The major sources of photometry were the $TychoII$ 
\citep{Hog2000}, $2MASS$ \citep{2MASS}, $WISE$ \citep{Wright2010}
catalogues, complemented with Str\"omgren \citep{Hauck1998}, Sloan
\citep{Brown2011} and GALEX (Galaxy Evolution
Explorer)\footnote{http://galex.stsci.edu/GR6/} photometry, when
available.  The second step consisted in estimating the extinction
value for each target. This task was accomplished by adopting the values from
\citet{Schlafly2011} for all the stars with two exceptions: HD\,43509 and HD\,185658 for which 
the \citet{Schlafly2011} value were too high for stars relatively close to the sun 
(i.e. $E(B-V)\sim0.7$ and $E(B-V)\sim0.15$ mag, respectively). 
In these cases we first left $VOSA$ free to fit also the reddening. Afterwards, we redetermined
this value {\it a posteriori} using the effective temperature and gravity
estimated spectroscopically to find the expected intrinsic $(B-V)$
color as tabulated by  e.g. \citet{Kenyon1995}. A simple comparison with
the observed  $(B-V)$ provided the adopted reddening value. The same
procedure was applied to the remaining five stars, allowing us  to verify
the reliability of the reddening values by \citet{Schlafly2011}
adopted for these objects. The $E(B-V)$ values for each target are reported in coloumn (6) of
Table~\ref{tabPhotometry}. 
An additional parameter needed to build the SEDs is the distance of
the star. As shown in coloumn (5) of Table~\ref{tabPhotometry}: only
three stars possess {\it Hipparcos} parallax \citep{leeuwen}, for the
remaining ones $VOSA$ assigned a reference distance of 10 pc\footnote{The reader
should note that this choice does not affect our determination of T$_{\rm eff}$ and
$\log g$ since we are reproducing the shape of the SED and not the actual luminosities.}. 
The observed SEDs for all the investigated objects are
shown as filled circles in Fig.~\ref{sed}.  Finally, $VOSA$ performed
a least-square fit to these SEDs  by adopting the ATLAS9 Kurucz ODFNEW
/NOVER models \citep{castelli97} to obtain a first estimate of 
effective temperature, gravity and metallicity (see Fig.~\ref{sed} and Table~\ref{tabPhotometry}). 
Note that  the uncertainties calculated by the VOSA package on these quantities, only
take into account for the internal error on SED fitting procedure,
resulting underestimated by at least a factor 2, mainly because of the
intrinsic limits of the method when, as in our case, mainly optical and infrared data is
available \citep[see e.g.][]{mcdonal12}. For this reason we have doubled the uncertainties 
on effective temperature, gravity and metallicity as listed in columns (8) to (10) of Table~\ref{tabPhotometry}. 
However, we stress again that obtaining precise values for the stellar parameters
and the relative uncertainties by means of the SED fitting procedure is not the purpose of the
present paper, as we only aim at  obtaining  quick and approximate starting points for the spectroscopic analysis. 

We also estimated the bolometric correction in the $V$-band ($BC_V$) for all our targets
with the aim to locate the stars in the HR diagram (see Section~\ref{HR}).
To calculate these values, we adopted the models by \citet{bcp98} where M$_{\rm bol,\odot}$\,=\,4.74 mag is assumed. 
We interpolated their model grids adopting the correct metal abundances as well as the values of $T_{\rm  eff}$ and 
$\log g$ derived spectroscopically (see next sections).

\begin{figure}                                       
\includegraphics[width=8.8cm]{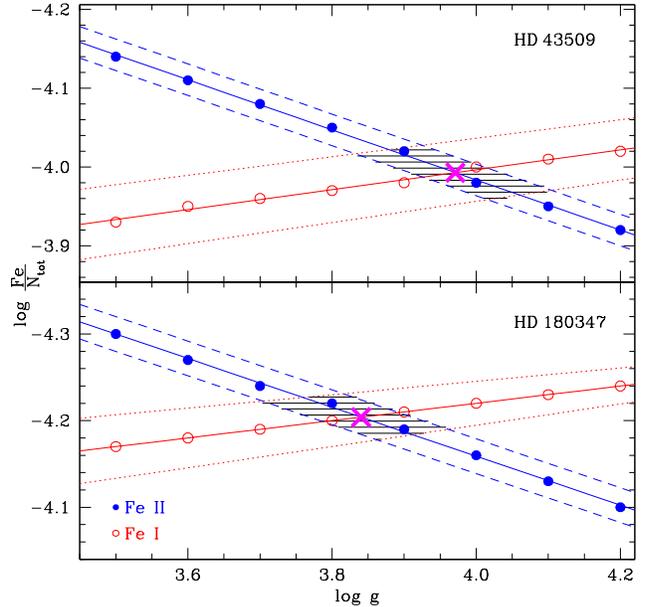}     
\caption{The ionization equilibrium of iron in the atmosphere of HD\,43509 anf HD\,180347. Open 
circles indicate abundances of Fe{\sc i} and filled circles those of Fe{\sc ii}, both plotted as 
a function of surface gravity. The dotted (for Fe{\sc i}) and dashed (for Fe{\sc ii}) lines are 
the 1~$\sigma$ bars, while the intersections at log\,$\approx$\,3.97 and log\,$\approx$\,3.85 are 
the adopted gravities.} 
\label{logg} 
\end{figure}

\begin{figure*}                                  
\includegraphics[width=16cm,bb= 18 144 592 450]{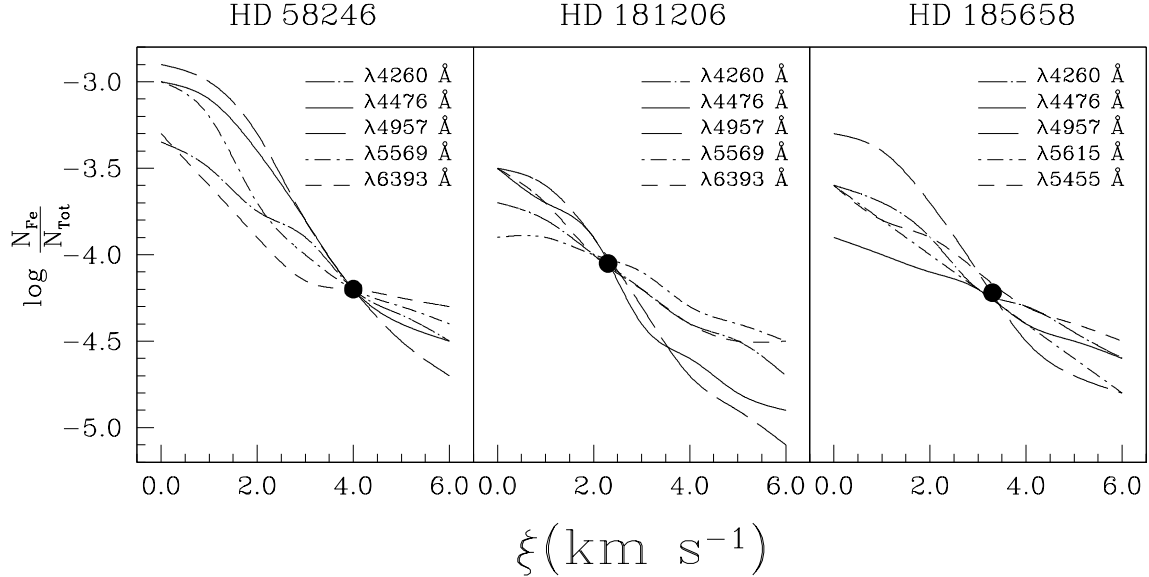}     
\caption{Blackwell diagrams plotted for Fe{\sc i} lines, left panel refers to HD\,58246, central panel to HD\,181206,
         and right panel to HD\,185658. The intersections of curves are the adopted microturbulence.} 
\label{black} 
\end{figure*} 

\section{Atmospheric parameters} 
\label{parameters_from_spectroscopy} 

In this section we present the spectroscopic analysis of our sample of
Am stars aimed at the determination of fundamental astrophysical quantities,
such as effective temperatures, surface gravities, rotational and
microturbulent velocities and chemical abundances.

The approach used in this paper has been succesfully used in other
papers devoted to this topics, see for instance
\citet{catanzaro14,catanzaro13,catanzaro12,catanzaro11}. In practice,
the procedure used for our targets was to minimize the difference
among observed and synthetic spectrum, using as goodness-of-fit
parameter the $\chi^2$ defined as

\begin{equation}
\chi^2\,=\,\frac{1}{N}\sum \left(\frac{I_{obs} - I_{th}}{\delta
    I_{obs}}\right) 
\end{equation}

\noindent
where N is the total number of points, I$_{obs}$ and I$_{th}$ are the
intensities of the observed and computed profiles, respectively, and
$\delta I_{obs}$ is the photon noise. Synthetic spectra were generated
in three steps: i) we computed LTE atmospheric models using the
ATLAS9 code \citep{kur93,kur93b}; ii) the stellar spectra were
synthesized using SYNTHE \citep{kur81}; iii) the spectra were
convolved for the instrumental and rotational broadening.

\begin{figure*}   
\begin{center}
\includegraphics[width=16.cm]{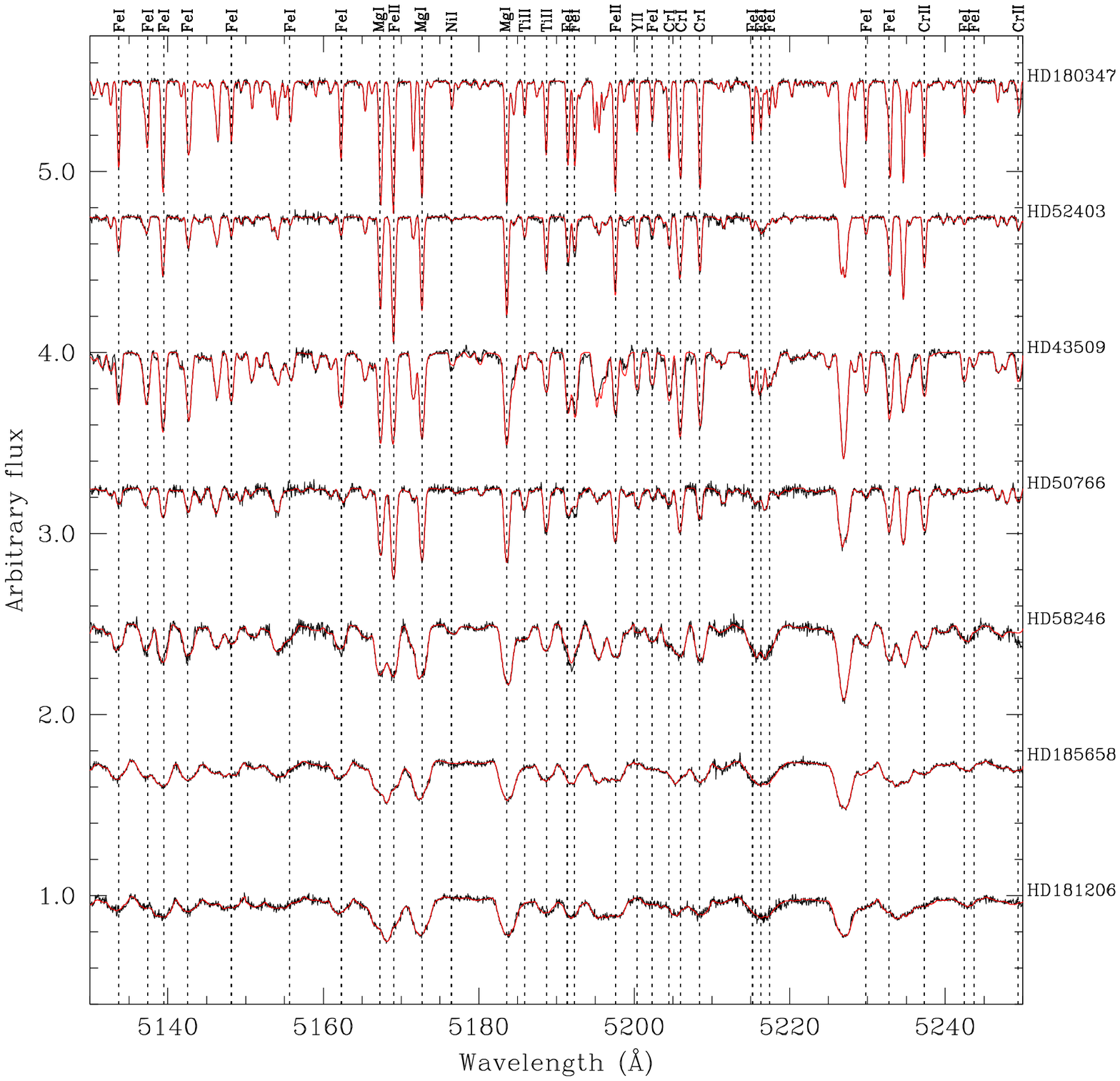}  
\caption{Example of our spectral synthesis in the range 5130 - 5250 {\AA}. Stars have been 
 ordered from top to bottom by increasing v\,$\sin i$.} 
\label{all} 
\end{center}   
\end{figure*}

We computed the $v \sin i$ of our targets by matching synthetic lines
profiles from SYNTHE to a significant number of metallic lines. To
this purpose, the spectral region around Mg{\sc i} triplet at $\lambda \lambda$5167-5183
{\AA} was particularly useful (see Fig.~\ref{all}). The resulting $v \sin i$ values are
listed in Table~\ref{param}.  

To determine stellar parameters as consistently as possible with the actual structure of the atmosphere,
we have extended \citet{leo97} and \citet{catanzaro04} iterative procedure to perform the abundances analysis:

\begin{description}

\item{(i)} $T_{\rm eff}$ was estimated by computing the ATLAS9 model
  atmosphere which produces the best match between the observed
  H$\alpha$ and H$\beta$ lines profile and those computed with SYNTHE.
  When $T_{\rm eff}$ is greater than 8000~K, Balmer lines become
  sensitive also to $\log g$. In these cases, we used another
  independent method to uniquely derive the temperature.  For the
  hottest stars of our sample, namely: HD\,50766, HD\,52403, and
  HD\,185658, we refined the temperatures by requiring that
  abundances obtained from differing ionization stages of iron must
  agree, in particular we used the Fe{\sc i} to Fe{\sc ii} line
  ratio As a first iteration, models were computed using solar
  Opacity Distribution Function (ODF) table and T$_{\rm eff}$ and
  $\log g$ derived in Sec.~\ref{parameters_from_sed}.

   The simultaneous fitting of these two lines (as shown in
   Fig.~\ref{balmer}) led to a final solution as the intersection of
   the two $\chi^2$ iso-surfaces. An important source of uncertainties
   arised from the difficulties in continuum normalization as it is
   always challenging for Balmer lines in echelle spectra.  We
   quantified the error introduced by the normalization to be at least
   100~K, that we summed in quadrature with the errors obtained by the
   fitting procedure. The final results for effective temperatures and
   their errors are reported in Table~\ref{param}.

   The surface gravities were estimated by three different methods
    according to the effective temperature of each object; for stars
    with T$_{\rm eff}$ greater than 8000~K (namely: HD\,50766,
    HD\,52433, HD\,58246, and HD\,185658), we derived $\log g$ by
    fitting the wings of Balmer lines. For stars cooler than this
    value, the wings of the Balmer lines loose their sensitivity to
    gravity, hence we adopted two alternative methods. For HD\,43509
    and HD\,180347 we exploited the ionization equilibrium Fe{\sc
      i}/Fe{\sc ii} (see Fig.~\ref{logg}), whereas for the fastest
    rotator of our sample, HD\,181206, the iron lines are too shallow
    and blended with other features. Therefore, for this star, we
    modeled the broad lines of the Mg{\sc i} triplet at
    $\lambda\lambda$~5167, 5172, and 5183 {\AA}, which are very
    sensitive to $\log g$ variations. In practice, we have first
    derived the magnesium abundance through the narrow Mg{\sc i} lines
    at $\lambda \lambda$~4571, 4703, 5528, 5711~{\AA} (not sensitive
    to $\log g$), and then we fitted the triplet lines by fine tuning
    the $\log g$ value. To accomplish this task is mandatory to take
    into account very accurate measurements of the atomic parameters
    of the transitions, i.e. $\log gf$ and the radiative, Stark and
    Van der Waals damping constants. Regarding $\log gf$, we used the
    values of \citet{aldenius07}, whereas the Van der Waals damping
    constant was taken from \citet{barklem00} ($\log \gamma_{\rm
      Waals}\,=\,-7.37$); the Stark damping constant is from
    \citet{fossati2011} ($\log \gamma_{\rm Stark}\,=\,-5.44$), and the
    radiative damping constant is from NIST database ($\log
    \gamma_{\rm rad}\,=\,7.99$).  For the sake of clarity, in
    Fig.~\ref{loggHD181206} (upper panel) we illustrated this
    procedure for the line Mg{\sc i} $\lambda$\,5183 {\AA}.  We
    computed seven synthetic spectral lines for different $\log g$,
    from 3.5 to 4.1~dex, with a step of 0.1~dex, keeping constant
    temperature and magnesium abundance. The adopted value of $\log g$
    was the one that minimized $\chi^2$ as shown in
    Fig.~\ref{loggHD181206} (bottom panel).

\begin{figure}                                       
\includegraphics[width=8.8cm]{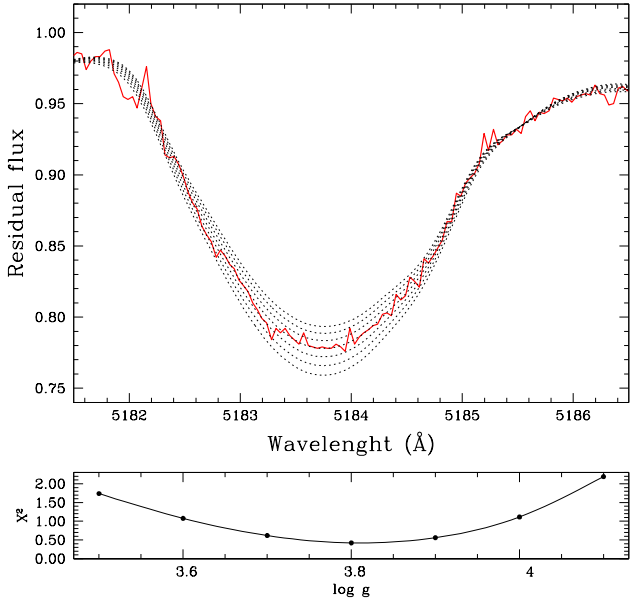}     
\caption{Example of fitting procedure to derive surface gravity
    for HD\,181206 by using Mg{\sc i} $\lambda$\,5183 {\AA}. {\it Upper panel}:
    comparison between observed (bold-red line) and synthetic lines
    (dotted) computed for seven different values of $\log g$, from 3.5
    to 4.1 dex (step 0.1 dex), with fixed temperature and
    abundance. {\it Bottom panel}: behavior of $\chi^2$ as a function
    of $\log g$.}
\label{loggHD181206} 
\end{figure}

\begin{figure*}   
\begin{center}
\includegraphics[width=12.cm,angle=-90]{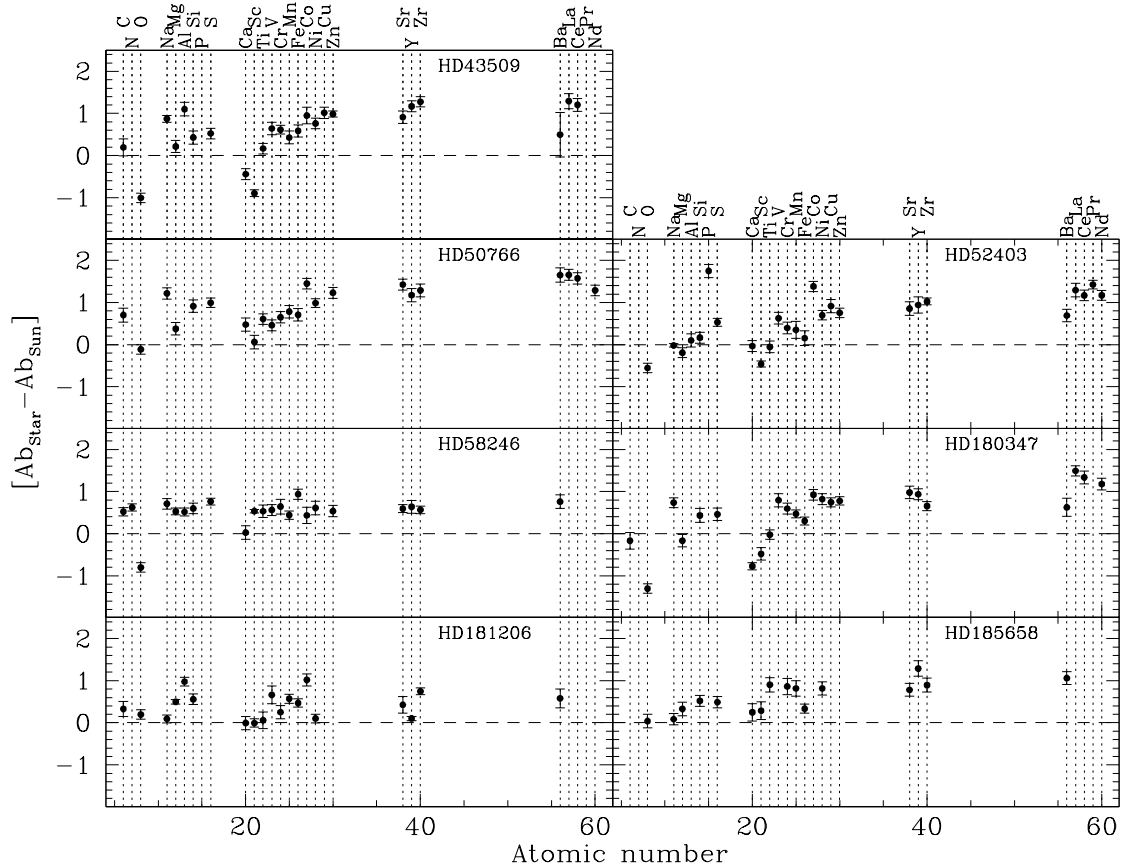}  
\caption{Chemical patterns derived for our targets. Horizontal dashed line corresponds 
         to solar abundance \citep{grevesse10}} 
\label{pattern} 
\end{center}   
\end{figure*} 

For estimating the microturbulence velocities, we applied two methods
according to the rotational velocities of our targets. For star with
$v\sin i \leq$~50~km~s$^{-1}$, $\xi$ was calculated by imposing that
the derived abundance does not depend on the equivalent widths
measured from a number of Fe{\sc i} lines for HD43509, HD\,50766,
HD\,52433, and HD\,180347. For the three stars with high rotational
velocities (HD\,58246, HD\,181206, and HD\,185658), since the great
rotational broadening makes difficult to find isolated lines, we built
the so-called Blackwell diagram. In practice, for a sample of Fe{\sc
  i} lines, we calculated the iron abundance for several values of
$\xi$, and plotted the result versus $\xi$. The loci in common with
all the lines gives the microturbulence value (see
Fig.~\ref{black}). Recently \citet{gebran14} published a study in
  which they discussed the microturbulence in a sample of more that
  170 A/F Am/Fm stars spread over a wide range of T$_{\rm eff}$ from
  6000~K to 10000~K.  From their studied it appears that values of
  $\xi \approx$ 5 km s$^{-1}$ are not uncommon in the range of
  temperatures typical of our stars. The adopted $\xi$ values for all the target
stars are listed in Table~\ref{param}.

Uncertainties in T$_{\rm eff}$, $\log g$, $\xi$ and $v \sin i$ as shown in Table~\ref{param} were
estimated by the change in parameter values which leads to an increases of $\chi^2$ by a unity \citep{lampton76}.

\item{(ii)} As a second step we determined the abundances of individual species by
  spectral synthesis. Therefore, we divided each of our spectra into
  several intervals, 50~{\AA} wide each, and derived the abundances in
  each interval by performing a $\chi^2$ minimization of the
  difference between the observed and synthetic spectrum. The
  minimization algorithm has been written in {\it IDL}\footnote{IDL (Interactive Data 
  Language) is a registered trademark of Exelis Visual Information Solutions} 
   language, using the {\it amoeba} routine. We adopted lists of spectral lines and
  atomic parameters from \citet{castelli04}, who updated the
  parameters listed originally by \citet{kur95}.

  \item{(iii)} if the metallicity obtained in the previous step,
    computed by averaging the abundances up to iron-peak elements, is
    different from the solar one, we repeated the calculations from
    scratch by adopting ODF close, as much as possible, to the derived
    metal abundances. See the papers by \citet{leo97,catanzaro04} on
    the importance of ODF in deriving abundances.

\end{description}

For each element, we calculated the uncertainty in the abundance to be
the standard deviation of the mean obtained from individual
determinations in each interval of the analyzed spectrum. For elements
whose lines occurred in one or two intervals only, the error in the
abundance was evaluated by varying the effective temperature and
gravity within their uncertainties given in Table~\ref{param}, $[
T_{\rm eff}\,\pm\, \delta T_{\rm eff}]$ and $[\log g\,\pm\,\delta \log
g]$, and computing the abundance for $T_{\rm eff}$ and $\log g$ values
in these ranges.  

As an example of our spectral synthesis, Fig.~\ref{all} shows the fit
obtained for our stars in the interval between 5130 and 5250 {\AA}. In that figure
stars have been ordered from top to bottom for increasing values of v\,$\sin i$.
\begin{table*}
 \centering
  \caption{Abundances inferred for the Am stars of our sample. Values are expressed in the usual form 
           $\log \frac{N_{\rm el}}{N_{\rm Tot}}$. Errors indicated in {\it italic} refer to those elements 
           for which spectral lines have been found in one or two intervals only (see text). The
           abundances reported for oxygen have been evaluated as discussed in Sect.~\ref{oxyg}.}
  %\begin{tabular}{l@{ }c@{ }c@{ }c@{ }c@{ }c@{ }c@{ }}
  \begin{tabular}{lccccccc}
  \hline
  \hline
   &         HD43509            &        HD50766              &      HD52403                 &       HD58246         &       HD180347                &          HD181206     &          HD185658     \\
\hline
C  & $-$3.41\,$\pm$\,0.20       &  $-$2.90\,$\pm$\,0.16       &           $--$               &  $-$3.08\,$\pm$\,0.08  &  $-$3.77\,$\pm$\,0.19        &  $-$3.28\,$\pm$\,0.17       &           $--$        \\
N  &          $--$              &           $--$              &           $--$               &  $-$3.58\,$\pm$\,0.07  &           $--$               &           $--$              &           $--$         \\
O  & $-$4.35\,$\pm$\,{\it 0.10} &  $-$3.45\,$\pm$\,{\it 0.10} &  $-$3.90\,$\pm$\,{\it 0.10}  &  $-$4.15\,$\pm$\,{\it 0.10}  &  $-$4.65\,$\pm$\,{\it 0.10}        &  $-$3.15\,$\pm$\,{\it 0.10}      &  $-$3.30\,$\pm$\,{\it 0.10}  \\
Na & $-$4.92\,$\pm$\,0.07       &  $-$4.58\,$\pm$\,0.13       &  $-$5.81\,$\pm$\,0.01        &  $-$5.08\,$\pm$\,0.11  &  $-$5.06\,$\pm$\,0.10        &  $-$5.70\,$\pm$\,0.08       &  $-$5.71\,$\pm$\,0.13  \\
Mg & $-$4.22\,$\pm$\,0.14       &  $-$4.06\,$\pm$\,0.14       &  $-$4.63\,$\pm$\,0.10        &  $-$3.91\,$\pm$\,0.07  &  $-$4.60\,$\pm$\,0.14        &  $-$3.94\,$\pm$\,0.03       &  $-$4.10\,$\pm$\,0.16  \\
Al & $-$4.48\,$\pm$\,0.15       &           $--$              &  $-$5.49\,$\pm$\,0.15        &  $-$5.07\,$\pm$\,0.09  &           $--$               &  $-$4.61\,$\pm$\,{\it 0.15} &           $--$        \\
Si & $-$4.09\,$\pm$\,0.15       &  $-$3.61\,$\pm$\,0.15       &  $-$4.36\,$\pm$\,0.12        &  $-$3.92\,$\pm$\,0.12  &  $-$4.09\,$\pm$\,0.16        &  $-$3.97\,$\pm$\,0.12       &  $-$4.00\,$\pm$\,0.13  \\
P  &          $--$              &           $--$              &  $-$4.87\,$\pm$\,{\it 0.15}  &           $--$         &           $--$               &           $--$              &           $--$         \\
S  & $-$4.39\,$\pm$\,0.12       &  $-$3.92\,$\pm$\,0.11       &  $-$4.38\,$\pm$\,0.08        &  $-$4.15\,$\pm$\,0.07  &  $-$4.46\,$\pm$\,0.14        &           $--$              &  $-$4.42\,$\pm$\,0.13  \\
Ca & $-$6.13\,$\pm$\,0.12       &  $-$5.22\,$\pm$\,0.15       &  $-$5.73\,$\pm$\,0.12        &  $-$5.67\,$\pm$\,0.15  &  $-$6.47\,$\pm$\,0.07        &  $-$5.70\,$\pm$\,0.15       &  $-$5.44\,$\pm$\,0.07  \\
Sc & $-$9.78\,$\pm$\,0.07       &  $-$8.82\,$\pm$\,0.15       &  $-$9.35\,$\pm$\,0.05        &  $-$8.35\,$\pm$\,0.10  &  $-$9.37\,$\pm$\,0.14        &  $-$8.89\,$\pm$\,0.09       &  $-$8.60\,$\pm$\,0.10  \\
Ti & $-$6.92\,$\pm$\,0.11       &  $-$6.50\,$\pm$\,0.11       &  $-$7.14\,$\pm$\,0.12        &  $-$6.55\,$\pm$\,0.13  &  $-$7.11\,$\pm$\,0.10        &  $-$7.02\,$\pm$\,0.18       &  $-$6.18\,$\pm$\,0.16  \\
V  & $-$7.46\,$\pm$\,0.13       &  $-$7.64\,$\pm$\,0.10       &  $-$7.48\,$\pm$\,0.11        &  $-$7.54\,$\pm$\,0.10  &  $-$7.31\,$\pm$\,0.13        &  $-$7.44\,$\pm$\,0.19       &           $--$         \\
Cr & $-$5.78\,$\pm$\,0.10       &  $-$5.74\,$\pm$\,0.13       &  $-$6.00\,$\pm$\,0.13        &  $-$5.75\,$\pm$\,0.17  &  $-$5.80\,$\pm$\,0.12        &  $-$6.14\,$\pm$\,0.14       &  $-$5.53\,$\pm$\,0.19  \\
Mn & $-$6.17\,$\pm$\,0.15       &  $-$5.82\,$\pm$\,0.14       &  $-$6.25\,$\pm$\,0.19        &  $-$6.16\,$\pm$\,0.09  &  $-$6.14\,$\pm$\,0.08        &  $-$6.03\,$\pm$\,0.09       &  $-$5.79\,$\pm$\,0.18  \\
Fe & $-$3.94\,$\pm$\,0.14       &  $-$3.83\,$\pm$\,0.15       &  $-$4.38\,$\pm$\,0.16        &  $-$3.60\,$\pm$\,0.11  &  $-$4.23\,$\pm$\,0.08        &  $-$4.06\,$\pm$\,0.08       &  $-$4.20\,$\pm$\,0.10  \\
Co & $-$6.00\,$\pm$\,0.15       &  $-$5.60\,$\pm$\,0.11       &  $-$5.66\,$\pm$\,0.09        &  $-$6.61\,$\pm$\,0.17  &  $-$6.12\,$\pm$\,0.10        &  $-$6.03\,$\pm$\,0.12       &           $--$         \\
Ni & $-$5.05\,$\pm$\,0.12       &  $-$4.82\,$\pm$\,0.09       &  $-$5.12\,$\pm$\,0.10        &  $-$5.20\,$\pm$\,0.15  &  $-$4.99\,$\pm$\,0.12        &  $-$5.71\,$\pm$\,0.09       &  $-$5.00\,$\pm$\,0.15  \\
Cu & $-$6.83\,$\pm$\,0.13       &           $--$              &  $-$6.93\,$\pm$\,{\it 0.15}  &           $--$         &  $-$7.10\,$\pm$\,{\it 0.15}  &           $--$              &           $--$         \\
Zn & $-$6.49\,$\pm$\,0.05       &  $-$6.25\,$\pm$\,{\it 0.12} &  $-$6.72\,$\pm$\,0.09        &  $-$6.93\,$\pm$\,0.12  &  $-$6.70\,$\pm$\,0.08        &           $--$              &           $--$         \\
Sr & $-$8.25\,$\pm$\,0.13       &  $-$7.74\,$\pm$\,0.10       &  $-$8.31\,$\pm$\,{\it 0.14}  &  $-$8.57\,$\pm$\,0.06  &  $-$8.19\,$\pm$\,0.12        &  $-$8.74\,$\pm$\,{\it 0.18} &  $-$8.38\,$\pm$\,0.13  \\
Y  & $-$8.66\,$\pm$\,0.12       &  $-$8.65\,$\pm$\,0.15       &  $-$8.88\,$\pm$\,0.18        &  $-$9.18\,$\pm$\,0.14  &  $-$8.89\,$\pm$\,0.12        &  $-$9.72\,$\pm$\,0.10       &  $-$8.53\,$\pm$\,0.18  \\
Zr & $-$8.18\,$\pm$\,0.11       &  $-$8.16\,$\pm$\,0.15       &  $-$8.43\,$\pm$\,0.07        &  $-$8.89\,$\pm$\,0.07  &  $-$8.80\,$\pm$\,0.09        &  $-$8.71\,$\pm$\,0.06       &  $-$8.56\,$\pm$\,0.16  \\
Ba & $-$9.36\,$\pm$\,0.51       &  $-$8.23\,$\pm$\,0.14       &  $-$9.16\,$\pm$\,0.12        &  $-$9.09\,$\pm$\,0.13  &  $-$9.23\,$\pm$\,0.19        &  $-$9.27\,$\pm$\,0.20       &  $-$8.80\,$\pm$\,0.12  \\
La & $-$9.64\,$\pm$\,0.17       &  $-$9.28\,$\pm$\,0.12       &  $-$9.64\,$\pm$\,0.15        &           $--$         &  $-$9.44\,$\pm$\,0.11        &           $--$              &           $--$         \\
Ce & $-$9.25\,$\pm$\,0.14       &  $-$8.88\,$\pm$\,0.13       &  $-$9.29\,$\pm$\,0.12        &           $--$         &  $-$9.12\,$\pm$\,0.14        &           $--$              &           $--$         \\
Pr &          $--$              &           $--$              &  $-$9.89\,$\pm$\,0.09        &           $--$         &          $--$                &           $--$              &           $--$         \\
Nd &          $--$              &  $-$9.32\,$\pm$\,0.12       &  $-$9.44\,$\pm$\,0.11        &           $--$         &  $-$9.44\,$\pm$\,0.13        &           $--$              &           $--$         \\
\hline
\end{tabular}
\label{abund}
\end{table*}

\section{Individual stars}
\label{singleStarAbundances}

In this section we present the results of the abundance analysis
obtained for each star in our sample.  The derived abundances and the
estimated uncertainties, expressed as $\log \frac{N_{el}}{N_{Tot}}$,
are reported in Table~\ref{abund}.  The abundance patterns for each
star, expressed in terms of solar values \citep{grevesse10}, are shown
in Fig.~\ref{pattern}.

\subsection{HD\,43509}
Classified as Am star by \citet{bidelman83}, this object was never observed
at high spectral resolution to date. Our study confirms the Am peculiarity of HD\,43509,
since we obtained calcium and scandium, respectively, $\approx$\,0.5
and $\approx$\,1.0 dex under the solar value, and iron-peak elements 
generally overabundant by 1.0~dex.

\subsection{HD\,50766}

The only previous spectral classification of this star is contained in
a paper published in late 70's by \citet{claunsen79}, in which the
object is classified as an Am star with spectral peculiarity
kA2mA7. No other analysis has been published to date.

Usually Am stars are defined as objects with enhanced iron-peak elements 
and solar or underabundance of calcium and/or scandium. In our analysis we
found overabundance of iron-peak and heavy elements such as: strontium, yttrium,
zirconium and barium (almost 2 dex larger than the corresponding solar
values), while scandium are almost normal in content. 

Thus, rather that Am star we can classify this object more properly as a marginal 
metallic star (Am:).

\subsection{HD\,52403}

This object was also observed by \citet{claunsen79}, which have
derived a spectral classification of kA2mA8. A recent spectral
classification A1{\sc iii}, given by \citet{abt04}, is in agreement
with our results for what concerns the value of $log
g$\,=\,3.5$\pm$0.1~dex, corresponding to a moderately evolved star. 
Our effective temperature is somewhat lower than that corresponding to a
spectral type A1 as derived by \citet{abt04}, but it is in good
agreement, at least within the errors, to the determination given by
\citet{mcdonal12}, who derived T$_{\rm eff}$\,=\,8216~K by modeling
its spectral energy distribution.

Our analysis shows clear evidence that this star could be considered
as a marginal Am star. In fact, we derived normal abundance of calcium
and light elements, an underabundance of $\approx$\,0.5~dex of
scandium, while iron-peak and heavy elements show overabundance up to
1 dex.

\subsection{HD\,58246}

HD\,58246 is the primary component of a visual double system
identified as an Am stars by \citet{olsen80} and later classified as
kA5hF2mF2 by \citet{abt08}. \citet{mcdonal12} derived T$_{\rm
  eff}$\,=\,8323~K by modeling its SED reconstructed by using several 
literature data; this value is in agreement with the one reported by us 
in Table~\ref{param}. 

The abundance analysis shows a slight overabundance of metals
 ($\approx$~0.5~dex), while calcium is solar in
content. This result is consistent with a classification of marginal
Am star.

\subsection{HD\,180347}

This star listed in the {\it Catalogue of Ap, HgMn and Am stars} \citep{renson09}, is one 
of the 15 Am stars present in the {\it Kepler} field of view. \citet{smalley11} observed 
pulsations in this object with a maximum amplitude less than 0.01 mmag. Later on, \citet{mcdonal12} 
deriving T$_{\rm eff}$\,=\,7685~K by modeling its SED, quite in agreement with our result.

Our analysis highlighted a chemical pattern typical of Am stars, with
underabundance of calcium ($\approx$\,0.7~dex) and scandium
($\approx$\,0.5~dex) and overabundances of iron-peak and heavy
elements. Carbon, oxygen, magnesium, and titanium display a solar-like abundance,
while some light elements as nitrogen, sodium, and phosphorus are clearly overabundant.  
Slight overabundances of silicon and sulfur ($\approx$\,0.5~dex) have been observed as well.

\subsection{HD\,181206}

Reported as Am star in the {\it Catalogue of Ap, HgMn and Am stars}
\citep{renson09} (spectral type A5 derived from K line),
HD\,181206 has been found pulsating in the superWASP survey by \citet{smalley11}. It is
also in the {\it Kepler} field of view. Recently this star has been investigated by
\citet{tkachenko12}, who found T$_{\rm eff}$\,=\,7478\,$\pm$\,41~K,
$\log g$\,=\,3.74\,$\pm$\,0.18~dex, and $\xi$\,=\,3.55$\pm$\,0.24~km s$^{-1}$.
By using the spectrum synthesis method described in their paper, they also obtained
[Fe/H]\,=\,$-$0.30\,$\pm$\,0.10~dex and a general trend of the chemical pattern
under the solar standard composition, with underabundance up to $-$0.45~dex
in the case of titanium. These chemical abundances are then not compatible to
those typical of Am stars. 

Besides T$_{\rm eff}$ and $\log g$ estimated in the present work are
very close or even consistent (in the case of gravity, within the
experimental errors), with those reported by \citet{tkachenko12}, our
results do not support their conclusions about the non-Am nature of
HD\,181206.  In fact, we found slight overabundances (from 0.5 up to
1~dex) of Fe and iron-peak elements, as vanadium, cobalt, and nickel,
as well as strontium, zirconium and barium. In particular we found
[Fe/H]\,=\,0.47\,$\pm$\,0.10~dex. Our calcium and scandium abundances
are instead solar in content.

Given the discrepancy between our results and those obtained by
\citet{tkachenko12}, we tried to derive the iron abundance using an
independent method. Elemental abundance of Fe was derived from the
measurements of line EWs using the 2013 version of MOOG
\citep{sneden73} and assuming LTE conditions. Model atmosphere was
computed with ATLAS9, by using the values of T$_{\rm eff}$, $\log g$,
and $\xi$ found by \citet{tkachenko12}.  With this parameters, we
obtained [Fe/H]\,=\,0.29\,$\pm$\,0.17~dex, which is consistent within
the errors with the results of our spectral synthesis.

Thus, our results confirm the marginal metallic (Am:) nature of this
object, in agreement with what reported in \citet{renson09}.

\subsection{HD\,185658}

Reported as Am star in the {\it Catalogue of Ap, HgMn and Am stars}
\citep{renson09}, this star was not detailed studied up to
date, at least to our knowledge.

From our work, we can classify this target as a marginal metallic star
(Am:), with calcium and scandium solar in content and overabundances
of iron-peak and heavy elements.

\section{Oxygen abundance}
\label{oxyg}
\begin{figure}   
\centering
\includegraphics[width=11.cm,angle=-90]{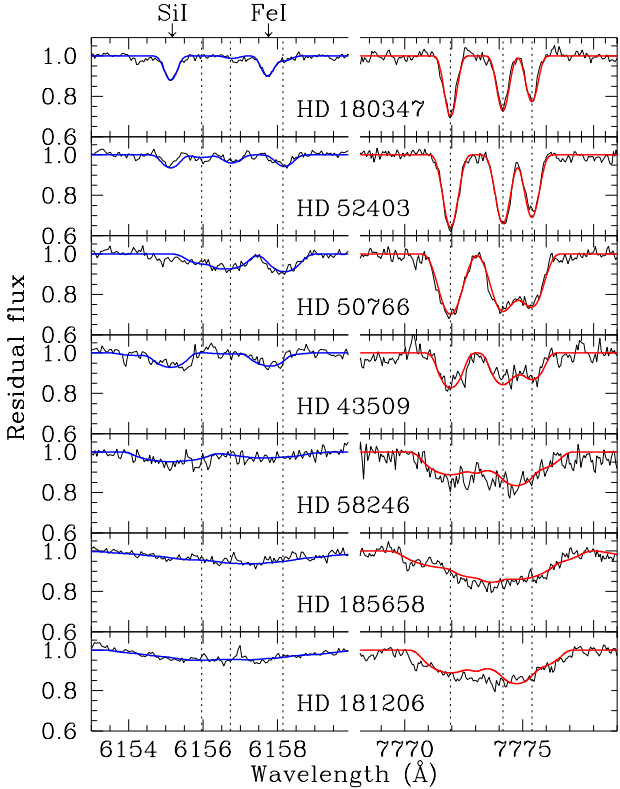}  
\caption{{\it Left side:} LTE synthetic spectra computed with
    ATLAS9+SYNTHE for O{\sc i} $\lambda$6155-8 {\AA} triplet (blue
    line) overimposed with the observations (black line). Arrows on
    top mark the positions of two lines possibly blended with oxygen,
    respectively Si{\sc i} $\lambda$6155.693 {\AA} and Fe{\sc i}
    $\lambda$6157.729 {\AA}. {\it Right side:} Non-LTE synthetic
    spectra computed with ATLAS9+SYNSPEC for the O{\sc i}
    $\lambda$7771-5 {\AA} triplet (red line) overimposed with the
    observations (black lines).  Dashed vertical lines mark the
    positions of triplets. Stars are ordered from top to bottom for
    increasing projected rotational velocities.} 
\label{synspec} 
\end{figure}

The oxygen abundance, one of the $\alpha$ elements, in the stellar
atmospheres is in general an important quantity in the scenario of the
Galactic evolution, as well as in the stellar structure and evolution
theories. With the exception of the infrared O{\sc i}\,$\lambda$7771-5~{\AA} 
triplet, the oxygen lines present in our
spectral range are weak and, in the case of rapid rotators, they are
blended with other lines or even confused with the continuum. For
instance, left side of Fig.~\ref{synspec}, show observed O{\sc i}
$\lambda$6155-8 {\AA} triplet (marked with dashed vertical lines)
with overimposed the LTE synthetic spectrum computed as described in
Sect.~\ref{parameters_from_spectroscopy}.  The IR triplet is
instead a strong feature easily measured also when high rotational
velocity makes difficult the detection of weak lines, see for
  clarity right side of Fig.~\ref{synspec}. Moreover, these lines lie
in a clean part of the spectrum (telluric lines are not present) and
they do not suffer any kind of blending, nor from atomic or molecular
lines. Despite their strength places them in the saturation part of
the curve-of-growth which, however, is not yet completely flat and
still provides a sufficient abundance sensitivity for these lines. All
these considerations make the O{\sc i} $\lambda$7771-5~{\AA} triplet a
good candidate to derive oxygen abundances in our targets. As to
Am stars, this triplet has already been used in literature to derive
oxygen abundance \citep[see for instance][]{van89,takeda97}, and the
general behavior is an underabundance with respect to the solar value.

\begin{figure}   
\includegraphics[width=8.8cm]{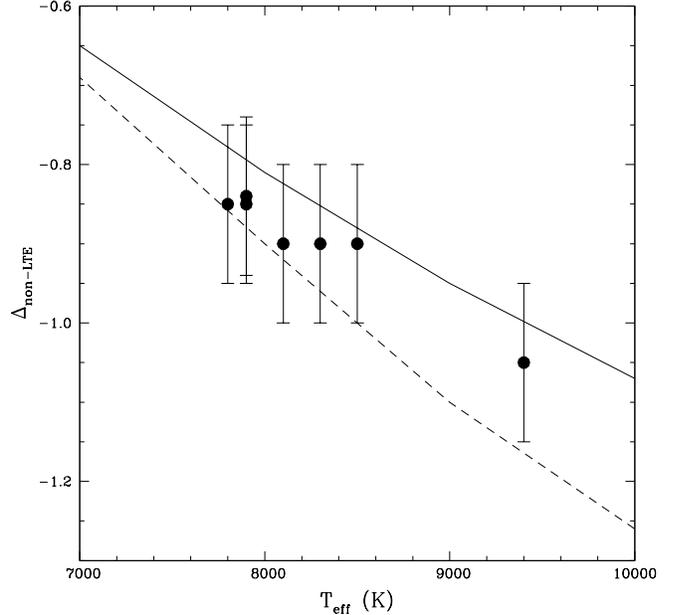}  
\caption{NLTE corrections for oxygen abundances computed for
    O{\sc i}$\lambda$7771-5~{\AA} triplets. Lines represents the
    theoretical corrections computed by \citet{sitnova13} with
    \citet{barklem07} collisional rate (solid line) and with these
    parameters scaled by a factor of 4 (dashed line).}
\label{nlte}
\end{figure}

Previously many authors noticed that this triplet gives systematically
higher abundance when compared to those obtained from other lines, in
some cases even by one order of magnitude.  One reason is that the IR
O{\sc i} lines are formed under conditions far from local
thermodynamic equilibrium (LTE). In fact, non-LTE spectral synthesis
leads to a strengthening of lines and to a decrease in the abundances
derived from these lines, as a natural consequence. In the case of
A-type main sequence stars, the usage of non-LTE does not solve the
problem, and we still have abundances from $\lambda$7771-5~{\AA}
triplet higher than those derived from visible lines.  This problem
has been investigated in the study by \citet{sitnova13}, who 
performed non-LTE calculations for this triplet by using model atom
by \citet{przybilla00}, modified according to the calculations of
cross sections for the excitation of O{\sc i} transitions under
collisions with electrons, performed by \citet{barklem07}. According
to their results, they suggested that the introduction of a scaling
factor of 1/4 to the rate coefficients for collisions with electrons
can reconcile the oxygen abundance derived from different lines.

For the stars of our sample, we explored the possibility of
  deriving the oxygen abundance by spectral synthesis of the IR
  triplet at $\lambda$7771-5~{\AA}.  We also compared the results
  obtained both with LTE and non-LTE approaches.  The LTE analysis has
  been performed as for the other elements by using ATLAS9+SYNTHE,
  while for what concerns non-LTE analysis we used ATLAS9+SYNSPEC
  \citep{hubeny00}. SYNSPEC reads the same input model atmosphere
  previously computed using ATLAS9 and solves the radiative transfer
  equation, wavelength by wavelength in a specified spectral
  range. SYNSPEC also reads the same Kurucz list of lines we used for
  the metal abundances. SYNSPEC permits to compute the line profiles
  considering an approximate non-LTE treatment, by means of the second-order 
  escape probability theory \citep[for details see the paper by][]{hubeny86}.

For each star, the oxygen abundance was calculated by fitting the observed 
triplet to the synthetic one computed as described before. Model atom has been 
taken from SYNSPEC web site\footnote{http://nova.astro.umd.edu/Synspec49/synspec-frames-atom.html}
The atomic parameters for the spectral lines were taken from VALD database \citep{kupka99}. 
The values that best reproduced the observed triplets are reported in Table~\ref{abund}, while the
  synthetic profiles are shown in Fig.~\ref{synspec}. As an additional
  check, we also compared in Fig.~\ref{nlte}, our non-LTE corrections,
  i.e. $\Delta_{\rm non-LTE} = \log \epsilon_{\rm nLTE} - \log
  \epsilon_{\rm LTE}$, with the theoretical predictions computed by
  \citet{sitnova13} for various effective temperature, $\log
  g$\,=\,4.0, solar chemical composition, oxygen abundance fixed to
  -3.21, and $\xi$\,=\,2~km~s$^{-1}$.  The behavior of our
  $\Delta_{\rm non-LTE}$ as a function of T$_{\rm eff}$ reflects the one
  computed by latter authors.

  In our sample of stars, oxygen appears to show sub-solar abundances
  up to $\approx$\,1~dex in the case of the two Am stars HD\,43509 and
  HD\,180347, moderate underabundances for HD\,50766, HD\,52403, and
  HD\,58246, while it is solar or slight overabundant for HD\,185658
  and HD\,181201, respectively.  This is in agreement with the
  previous literature results. 

\begin{figure}    
\centering                               
\includegraphics[width=9.5cm]{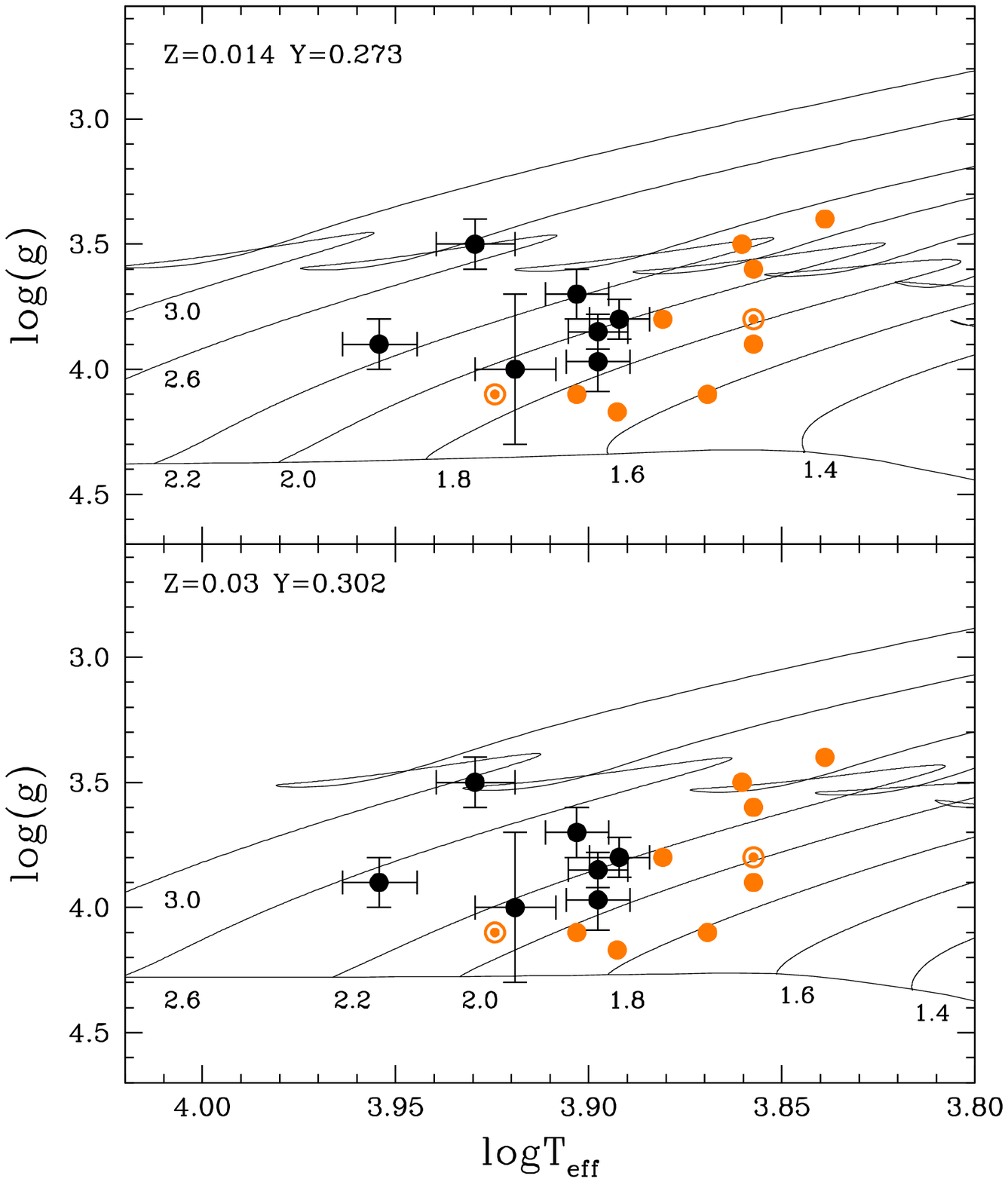} 
\caption{Both top and bottom panels show the $\log T_{\rm eff}$-$logg$
  diagram for the seven stars investigated in this paper (black filled
  circles). Similarly, the light orange symbols (note that filled-empty
  circles show non-Am stars) show the location of the stars studies in
  our previous works \citep[][]{balona11,catanzaro13,catanzaro14}.
  Top and bottom panels show also the evolutionary tracks (thin solid
  lines) for the labeled masses as well as the ZAMS from the {\it
    PARSEC} database for Z=0.014 Y=0.273 and Z=0.03 Y=0.302,
  respectively.}
\label{hrg} 
\end{figure}

\section{Position in the HR Diagram}
\label{HR}

\begin{figure}    
\centering                               
\includegraphics[width=9.5cm]{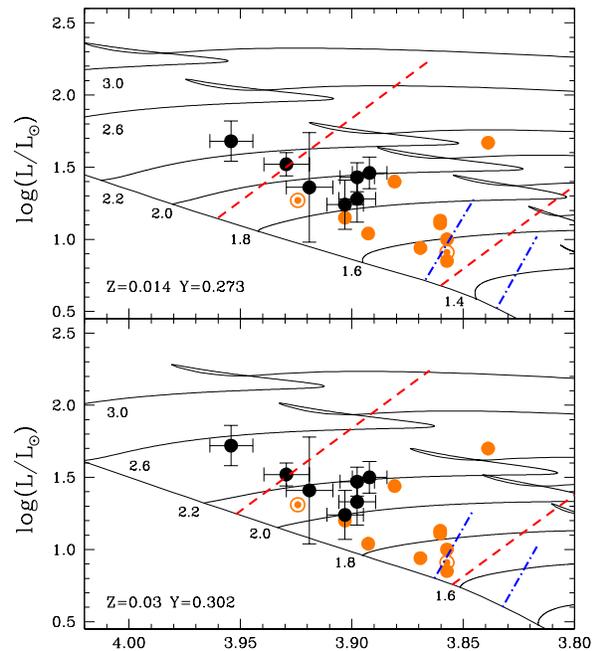} 
\caption{Top and bottom panels show the HR diagram for the seven stars
  investigated in this paper (black filled circles).  Similarly, the
  light orange symbols (note that filled-empty circles show non-Am
  stars) show the location of the stars studies in our previous works
  \citep[][]{balona11,catanzaro13,catanzaro14}.  For comparison
  purposes we overplot the $\delta$\,Sct (red dashed lines) and the
  theoretical edges of the $\gamma$\,Dor (blue dotted-dashed lines)
  instability strips by \citet{breger98} and \citet{guzik},
  respectively. Top and bottom panels show also the evolutionary
  tracks (thin solid lines) for the labeled masses as well as the
  ZAMS from the {\it PARSEC} database for Z=0.014 Y=0.273 and Z=0.03
  Y=0.302, respectively.}
\label{figHR} 
\end{figure} 

An accurate location of the Am star in the HR
diagram is useful to investigate possible systematic differences in
the region occupied by these objects with respect to normal A stars as 
well as to properly constrain the pulsation instability strip when
these stars are pulsating as $\delta$\,Sct and/or $\gamma$\,Dor
variables \citep[see][]{smalley11,catanzaro14}.

To locate our target stars on the HR diagram we have first to
  estimate the value of $\log L/L_{\odot}$ for each target. In
  particular, for the two stars whose parallaxes were measured by
  Hipparcos (namely, HD\,52403 and HD\,58246) the luminosities can be
  easily calculated on the basis of the data listed in
  Table~\ref{tabPhotometry}. As for the remaining five stars, we have
  first to link the measured quantities, $T_{\rm eff}$ and $\log g$ to
  the unknown value of $\log L/L_{\odot}$. To this aim, we can use the
  recent evolutionary tracks available from the {\it PARSEC} (Padova
  Trieste Stellar Evolution Code) database \citep[see e.g.][and the
  web site http://stev.oapd.inaf.it/cgi-bin/cmd]{Bressan2012},
  provided that we know the chemical composition of the target stars
  (Z and Y). 
  In this context, we note that all the Am stars analyzed here show  moderate to high supersolar
  $[M/H]$\footnote{$[M/H]=\log(N_M/N_H)-\log(N_M/N_H)_{\odot}$, where
    $N_M$ and $N_H$ are the number of metal and hydrogen atoms per
    unit of volume, respectively. In turn, $[M/H]$ is related to Z by means of
    the equation $\log$Z=$[M/H]$-$\log$Z$_{\odot}$.} values,
  ranging from $\sim0.2$ to $\sim0.6$ dex (with a large scatter
  $\sim$0.2-0.4 dex,  see also Fig.~\ref{pattern} and
  Table~\ref{abund}).
However, estimating the actual value of Z for an 
  Am star is not an easy task, given that the atmospheric abundances (the only
  ones we can measure) are significantly altered by the still not well
  understood physical processes that are at the base of the Am
  phenomenon. This occurrence makes almost impossible to infer the precise
  value of Z to be used for the investigated stars. Thus, we decided
  to adopt evolutionary tracks with
both solar and twice solar abundance values so that to take into
account the uncertainty on the actual metallicity of the targets. 
More in detail, we used the {\it PARSEC} version v1.2s tracks 
for Z=0.014 Y=0.273\footnote{Note that {\it PARSEC} models adopt
  Z$_{\odot}$=0.0152 \citep[for details see Sect. 2.1
  in][]{Bressan2012}, but the tracks are available for Z=0.014} and  Z=0.03  Y=0.302. 
To test  the  sensitivity of the results to evolutionary models
calculated with  different chemical composition, we show in 
  Fig.~\ref{hrg} the $\log T_{\rm eff}$-$logg$ diagram for the
  seven stars investigated in this paper (as well as the Am stars
  investigated in our previous papers, see the caption of the figure) in comparison with the
  evolutionary models quoted above. As a result, it appears that the
  masses of the targets inferred from the evolutionary tracks are on
  average about 0.2 $M_{\odot}$ larger if Z=0.014 is assumed, being
however  the uncertainty due to effective temperature and gravity of the
same order of magnitude. 
Passing to the estimate of $\log L/L_{\odot}$ as a function of $T_{\rm eff}$ and $\log
  g$ based on evolutionary models, we first selected tracks with masses ranging from 1.2 to
  4.8 $M_{\odot}$ and then did not consider evolutionary phases later than
  the base of the red giant branch. A simple least-square fit among
  these models gives the following equations for Z=0.014 and Z=0.03, respectively:

\begin{eqnarray}
\label{eq}
\log L/{\rm L}_\odot = -(14.909\,\pm\,0.006)+(5.406\,\pm\,0.002) \log T_{\rm eff}\\
 -(1.229\,\pm\,0.001) \log g~~{\rm rms=0.042~dex.} \nonumber 
\end{eqnarray}

\begin{eqnarray}
\label{eq1}
\log L/{\rm L}_\odot = -(14.814\,\pm\,0.006)+(5.380\,\pm\,0.001) \log T_{\rm eff}\\
 -(1.215\,\pm\,0.001) \log g~~{\rm rms=0.038~dex.} \nonumber 
\end{eqnarray}

Finally, we used the Eq.~\ref{eq} and~\ref{eq1} in conjunction with the observed
$T_{\rm eff}$ and $\log g$ to estimate the $\log L/{\rm L}_\odot$ for
the five targets without Hipparcos parallaxes.  The result of the
whole procedure is shown in Table~\ref{tabPhotometry} columns (11)
and (12), as well as in Fig.~\ref{figHR}, where we added the 11
stars analyzed in our previous works on Am stars \citep{catanzaro13,catanzaro14}.\footnote{Note that for
  homogeneity purposes, Eq.~\ref{eq} was used to recalculate the
  luminosity for stars HD\,113878, HD\,176843, HD\,179458 and
  HD\,187254.  As for the remaining seven stars, their luminosities
  were estimated on the basis of the Hipparcos parallaxes and hence
  reported here without changes.} The uncertainties on the
luminosities were calculated taking into account both the errors on
effective temperature and gravity as well as the rms of each
relationship. A comparison of col. (11) and (12)
in Table~\ref{tabPhotometry} reveals that the use of Eq.~\ref{eq} or
~\ref{eq1} to estimate the luminosity of the target stars do not
produce significant changes.

We also show in Fig.~\ref{figHR} for comparison purposes the edges of
the $\delta$\,Sct \citep[after][]{breger98} and $\gamma$\,Dor
\citep[after][]{guzik} instability strips, respectively. In passing,
we note that among the seven Am stars presented here, only HD\,181206
shows $\delta$\,Sct pulsations (plus rotational modulation) in the
{\it Kepler} light curve, whereas HD\,180347 and HD\,185658 show
rotational modulation \citep[see][]{Balona15}.  For the remaining
stars there is no information regarding their variability except for
HD\,43509 that was found constant by \citet{Paunzen2013}.

\section{Discussion and conclusion}

In this work we presented a spectroscopic analysis of a sample of
seven stars classified in literature as metallic Am stars. The analysis 
is based on high resolution spectra obtained at the 91-cm telescope of the 
Catania Astrophysical Observatory equipped with the new {\it CAOS} spectrograph. 
For each spectra we obtained fundamental parameters such as effective temperatures,
gravities, rotational and microturbulent velocities, and we performed
a detailed computation of the chemical pattern, as well. To overcome
the severe blending of spectral lines, caused by the high rotational velocities of 
most or our targets, we applied the spectral synthesis method by using 
SYNTHE \citep{kur81} and ATLAS9 \citep{kur93} codes. We used also SYNSPEC 
\citep{hubeny00} to compute non-LTE abundances of oxygen from the IR 
triplet at $\lambda$7771-5~{\AA}. With the exception of HD\,181206, for which
another analysis exists in literature, the other stars were analyzed in detail 
for the first time in this study.

The values of T$_{\rm eff}$ and $\log g$ derived here have been used
to determine the luminosity of the stars and to place them on the HR
diagram.

Our analysis show that two stars are definitively Am stars, namely: HD\,43509
and HD\,180347, while HD\,50766, HD\,52403, HD\,58246, HD\,181206, and
HD\,185658 are marginal Am stars (Am:). For all the targets, the oxygen abundance 
has been derived separately by a non-LTE approach, modeling the triplet at 
$\lambda$7771-5~{\AA}. As a general results we found underabundances of oxygen
with the exception of HD\,185658 (solar) and HD\,181206 (+0.2~dex).

The availability of an instrument capable of high-resolution spectroscopy 
in a wide spectral range such as CAOS will allows us in the future to continue the
study of all the Am stars observed by the {\it K2}
extension of the {\it Kepler} mission that are visible from Catania Astrophysical 
Observatory and having V$\leq$10 mag. In particular, we will
be able to further clean the sample of Am stars observed in the  {\it K2} 
fields from non-Am objects, as well as to fully characterize the 
spectral properties of the truly Am objects.

\section*{Acknowledgments} 
The authors wish to thank Dr. L. Balona and Dr. P. G. Prada Moroni for helpful discussions.\\
This publication makes use of VOSA, developed under the Spanish Virtual Observatory 
project supported from the Spanish MICINN through grant AyA2011-24052.\\
This research has made use of the SIMBAD database and VizieR catalogue access tool,
operated at CDS, Strasbourg, France.\\
This publication makes use of data products from the Wide-field
Infrared Survey Explorer, which is a joint project of the University of
California, Los Angeles, and the Jet Propulsion Laboratory/California
Institute of Technology, funded by the National Aeronautics and Space
Administration. \\
Atomic data compiled in the DREAM data base (E. Biemont, P. Palmeri \& P. Quinet, Astrophys. Space Sci. 269-270, 635 (1999))
were extracted via VALD (Kupka et al., 1999, A\&AS 138, 119, and references therein)

\label{lastpage} 
 
\end{document}